\def\simgt{\lower.5ex\hbox{$\; \buildrel > \over \sim \;$}}
\def\simlt{\lower.5ex\hbox{$\; \buildrel < \over \sim \;$}}

\documentclass[useAMS,usenatbib]{mn2e}

\usepackage{graphicx}

\title[The kinematics of edge-on stellar discs]
{Structure and kinematics of edge-on galaxy discs --
IV. The kinematics of the stellar discs.}
\author[M. Kregel \& P.~C. van der Kruit]
  {M. Kregel$^1$ and 
  P.~C. van~der~Kruit$^1$\thanks{E-mail: vdkruit@astro.rug.nl}\\
  $^1$Kapteyn Astronomical Institute, University of Groningen,
  P.O.Box 800, 9700AV Groningen, the Netherlands}

\begin{document}

\date{Accepted. Received.}

\pagerange{\pageref{firstpage}--\pageref{lastpage}} \pubyear{2004}

\label{firstpage}

\maketitle

\begin{abstract}
The stellar disc kinematics in a sample of fifteen intermediate- to 
late-type  edge-on spiral galaxies are studied using
a dynamical modeling  technique. The sample covers a substantial range in
  maximum rotation velocity and deprojected face-on surface
  brightness and contains seven spirals with either a boxy- or
  peanut-shaped bulge. Dynamical models of the stellar discs are
constructed using the disc structure from 
$I$-band surface photometry and rotation curves observed in the gas.
The differences in the line-of-sight stellar kinematics between the models
and absorption line spectroscopy are minimized using
a least-squares approach. The modeling constrains the disc surface 
density and stellar radial velocity dispersion at a fiducial radius
through the free parameter 
$\sqrt{M/L}$ $(\sigma_{\rm z}/\sigma_{\rm R})^{-1}$, where
$\sigma_{\rm z}/\sigma_{\rm R}$ is the ratio of vertical and
radial velocity dispersion and  $M/L$ the disc mass-to-light ratio. 
For thirteen spirals a transparent model
  provides a good match to the mean line-of-sight
stellar velocity dispersion. Models that
  include a realistic radiative transfer prescription confirm that the
effect of dust on the observable stellar kinematics is small at the 
observed slit  positions. We discuss possible sources of
systematic error and conclude that most of these are likely to be small.
The exception is the neglect of the dark halo gravity, which has probably 
caused an overestimate of the surface density in the case of low
surface brightness discs.

This paper has been accepted for publication by MNRAS 
and is available in pdf-format
at the following URL:\\

http://www.astro.rug.nl/$\sim $vdkruit/jea3/homepage/paperIV.pdf.

\end{abstract}


\end{document}